# V-SMART-Join: A Scalable MapReduce Framework for All-Pair Similarity Joins of Multisets and Vectors

Ahmed Metwally
Google, Inc.
Mountain View, CA, USA
metwally@google.com

Christos Faloutsos
SCS, Carnegie Mellon University
Pittsburgh, PA, USA
christos@cs.cmu.edu

## ABSTRACT

This work proposes *V-SMART-Join*, a scalable MapReduce-based framework for discovering all pairs of similar entities. The *V-SMART-Join* framework is applicable to sets, multisets, and vectors. *V-SMART-Join* is motivated by the observed skew in the underlying distributions of Internet traffic, and is a family of 2-stage algorithms, where the first stage computes and joins the partial results, and the second stage computes the similarity exactly for all candidate pairs. The *V-SMART-Join* algorithms are very efficient and scalable in the number of entities, as well as their cardinalities. They were up to 30 times faster than the state of the art algorithm, *VCL*, when compared on a real dataset of a small size. We also established the scalability of the proposed algorithms by running them on a dataset of a realistic size, on which *VCL* never succeeded to finish. Experiments were run using real datasets of IPs and cookies, where each IP is represented as a multiset of cookies, and the goal is to discover similar IPs to identify Internet proxies.

## 1. INTRODUCTION

The recent proliferation of social networks, mobile applications and online services increased the rate of data gathering. Such services gave birth to Internet-traffic-scale problems that mandate new scalable solutions. Each online surfer contributes to the Internet traffic. Internet-traffic-scale problems pose a scalability gap between what the data analysis algorithms can do and what they should do. The MapReduce [11] framework is one major shift in the programming paradigms proposed to fill this gap by distributing algorithms across multiple machines.

This work proposes the *V-SMART-Join* (Versatile Scalable MApReduce all-pair similariTy Join) framework as a scalable exact solution to a very timely problem, all-pair similarity joins of sets, multisets and vectors. This problem has attracted much attention recently [2, 3, 4, 5, 6, 9, 10, 13, 22, 29, 33, 34] in the context of several applications. The applications include clustering documents and web content [3, 13, 34], detecting attacks from colluding attackers [22], refining queries and doing collaborative filtering [4], cleaning data [2, 10], and suggesting friends in social services based on common interests [12].

The motivating application behind this work is community discovery, where the goal is to discover strongly connected sets of entities in a huge space of sparsely-connected entities. The mainstream work in the field of community discovery [20, 27, 30, 36] has assumed the relationships between the entities are known *a priori*, and has proposed clustering algorithms to discover communities. While the relationships between entities are usually volunteered by domain experts, like in the case of bioinformatics, or by the entities themselves, like in social networks, this is not always the case. When information about the relationships is missing, it is reasonable to interpret high similarity between any two entities as an evidence of an existing relationship between them. Hence, our focus is to discover similar pairs of entities.

We propose using community discovery for classifying IP addresses as load balancing proxies. An Internet Service Provider (ISP) that assigns dynamic IP addresses (IPs for short) to its customers sends their traffic to the rest of the Internet via a set of proxy IPs. For advertisement targeting, and traffic anomalies detection purposes, it is crucial to identify these load balancing proxies, and treat each set of load balancers as one indivisible source of traffic. For instance, for the application of traffic anomalies detection based on the source IP of the traffic [23, 24], the same whitelisting/blacklisting decision should be taken for all the IPs of an ISP load balancer. For the application of targeting advertisement, the IP of the surfer gets resolved to a specific country or city, and the ads are geographically targeted accordingly. Some ISPs provide services in multiple locations, and their IPs span an area wider than the targeting granularity. No ads should be geo-targeted for the IPs of the same load balancer if the IPs resolve to multiple locations.

To that end, we propose representing each IP using a multiset, also known as a bag, of the cookies that appear with it, where the multiplicity of the cookies is the number of times it appeared with the IP. Identifying IPs of a load balancer reduces to finding all pairs of IPs with similar multisets of cookies. Representing IPs as multisets, as opposed to sets, makes the results more sensitive to the activities of the cookies, and hence increases the confidence in the results. A post-processing step is to cluster these IPs, where each pair of similar IPs are connected by an edge in an IP-similarity graph. A clusters correspond to IPs of the same load balancer. This work complements the work in [24] that estimates the number of users behind IPs, which can also be used for identifying large Internet proxies.

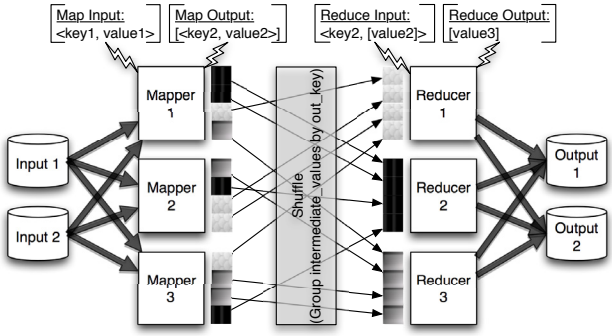

Figure 1: The MapReduce framework.

To discover all pairs of similar IPs, this work proposes *V-SMART-Join*, a scalable MapReduce based framework. The contributions of this work is as follows.

1. *Versatility*: *V-SMART-Join* is carefully engineered to work on vectors, sets, and multisets using a wide variety of similarity measures.

2. *Speed and Scalability*: *V-SMART-Join* employs a two stage approach, which achieves significant scalability in the number of entities, as well as their cardinalities, since it does not entail loading whole entities into the main memory. Moreover, *V-SMART-Join* carefully handles skewed data distributions.

3. *Wide Adoption*: The proposed *V-SMART-Join* algorithms can be executed on the publicly available version of MapReduce, Hadoop [1].

4. *Experimental Verification*: On real datasets, the *V-SMART-Join* algorithms ran up to 30 times faster than the state of the art algorithm, *VCL* [33].

The rest of the paper is organized as follows. The MapReduce framework is explained in § 2. In § 3, the problem is formalized and an insight is presented to build distributed algorithms. This insight is based on a classification of the partial results necessary to calculate similarity. The *V-SMART-Join* framework is presented in § 4. The *V-SMART-Join* algorithms are presented in § 5. The related work is reviewed in § 6. The experimental evaluation is reported in § 7, and we conclude in § 8.

## 2. THE MAPREDUCE FRAMEWORK

The MapReduce framework was introduced in [11] to facilitate crunching huge datasets on shared-nothing clusters of commodity machines. The framework tweaks the *map* and *reduce* primitives widely used in functional programming and applies them in a distributed computing setting.

Each record in the input dataset is represented as a tuple $\langle key_1, value_1 \rangle$. The first stage is to partition the input dataset, typically stored in a distributed file system, such as GFS[14], among the machines that execute the map functionality, the *mappers*. In the second stage, each mapper applies the map function on each single record to produce a list on the form $(\langle key_2, value_2 \rangle)^*$, where $(.)^*$ represents lists of length zero or more. The third stage is to shuffle the output of the mappers into the machines that execute the reduce functionality, the *reducers*. This is done by grouping the mappers' output by the key, and producing a reduce_value_list of all the $value_2$'s sharing the same value of $key_2$. In addition to $key_2$, the mapper can optionally output tuples by a secondary key. Each reducer would then receive the reduce_value_list sorted by the secondary key. Secondary keys are not supported by the publicly available version of MapReduce, Hadoop [1][1]. The input to the reducer is typically tuples on the form $\langle key_2, (value_2)^* \rangle$. For notational purposes, the reduce_value_list of key $k$ is denoted reduce_value_list$_k$. In the fifth stage, each reducer applies the reduce function on the $\langle key_2, (value_2)^* \rangle$ tuple to produce a list of values, $(value_3)^*$. Finally, the output of the reducers is written to the distributed file system. The framework is depicted in Figure 1.

MapReduce became the *de facto* distributed paradigm for processing huge datasets because it disburdens the programmer of details like partitioning the input dataset, scheduling the program across machines, handling failures, and managing inter-machine communication. Only the map and reduce functions on the forms below need to be implemented.

map:
$\langle key_1, value_1 \rangle \rightarrow (\langle key_2, value_2 \rangle)^*$

reduce:
$\langle key_2, (value_2)^* \rangle \rightarrow (value_3)^*$

For better fault tolerance, the map and reduce functions are required to be pure and deterministic. For higher efficiency, the same machines used for storing the input can be used as mappers to reduce the network load. In addition, partial reducing can happen at the mappers, which is known as *combining*. The *combine* function is typically the same as the reduce function. While combining does not increase the power of the framework, it reduces the network load[2].

The amount of information that need to fit in the memory of each machine is a function of the algorithm and the input and output tuples. In terms of the input and output tuples, during the map stage, at any time, the memory needs to accommodate one instance of each of the tuples $\langle key_1, value_1 \rangle$ and $\langle key_2, value_2 \rangle$. Similarly, during the reduce stage, the memory needs to accommodate one instance of each of $key_2$, $value_2$ and $value_3$. Nevertheless, accommodating multiple values of $\langle key_1, value_1 \rangle$, $\langle key_2, value_2 \rangle$ or $value_3$ allows for I/O buffering. Accommodating the entire reduce_value_list in memory allows for in-memory reduction.

For more flexibility, the MapReduce framework also allows for loading external data both when mapping and reducing. However, to preserve the determinism and purity of the map and reduce functions, loading is allowed only at the beginning of each stage. Moreover, the types of $key_1$, $key_2$, $value_1$, $value_2$ and $value_3$ are independent[3].

---
[1]Two ways to support secondary keys were proposed in [21]. One of them is not scalable, since it entails loading the entire reduce_value_list in the memory of the reducer, and the second solution entails rewriting the partitioner, the MapReduce component that assigns instances of $key_2$ to reducers. The second solution was adopted on the web page of [1]. We propose algorithms that avoid this limitation

[2]Combiners can be either dedicated functions or part of the map functions. A dedicated combiner operates on the output of the mapper. Dedicated combiners involve instantiation and destruction. On the other hand, an on-mapper-combiner is part of the mapper, is lightweight, but may involve fitting all the keys the mapper observes in memory, which can result in thrashing. This is discussed in details in [21]. We used dedicated combiners for higher scalability.

This framework, albeit simple, is powerful enough to serve as the foundation for an array of platforms. Examples include systems that support issuing SQL(-like) queries that get translated to MapReduce primitives and get executed in a distributed environment [25, 35, 32]. Another relevant example is adapting stream analysis algorithms to a distributed setting by the Sawzall system [26].

It is difficult to analyze the complexity of a MapReduce-based algorithm due to several factors, including the overlap between mappers, shufflers and reducers, the use of combiners, the high I/O and communication cost as compared to the processing cost. However, to the best of our abilities, we will try to identify the bottlenecks throughout the sequel.

Having described the necessary background, the insight for scalable MapReduce-based algorithms is described next.

## 3. PROBLEM FORMALIZATION AND INSIGHTS

We start by the formalization, and then use it to present the insight for more scalable solutions.

### 3.1 Formalizing the Problem

Given a set, $S$, of multisets, $M_1, \ldots, M_{|S|}$ on the alphabet $A = a_1, \ldots, a_{|A|}$, find all pairs of multisets, $\langle M_i, M_j \rangle$, such that their similarity, $Sim(M_i, M_j)$ exceeds some threshold, $t$. The similarity measure, $Sim(.,.)$ is assumed to be commutative. A multiset, identified by $M_i$, is represented as $M_i = \langle A, A \to \mathbf{N} \rangle = \{m_{i,1}, \ldots, m_{i,|A|}\}$, where $m_{i,k}$ represents the element in multiset $M_i$ that have the alphabet element $a_k$. More formally, $m_{i,k} = \langle a_k, f_{i,k} \rangle$ and $f_{i,k} \in \mathbf{N}$ is the *multiplicity* of $a_k$ in $M_i$. The cardinality of $M_i$ is denoted $|M_i| = \sum_{1 \le k \le |A|} f_{i,k}$. The set of alphabet elements that are present in $M_i$ is called its underlying set, $U(M_i)$. That is, $U(M_i) = a_k : f_{i,k} > 0$. Hence, $U(M_i) = \langle A, A \to \{0,1\} \rangle$. The underlying cardinality of $M_i$ is the number of unique elements present in $M_i$, i.e., $|U(M_i)| = |a_k : f_{i,k} > 0|$ [31]. The frequency of an element, $a_k$, denoted $Freq(a_k)$, is the number of multisets $a_k$ belongs to.

Representing multisets as non-negative vectors is trivial if $A$ is totally ordered. The semantics of sets can also be used to represent the more general notion of multisets. A multiset can be represented as a set by expanding each element $m_{i,k}$ into the elements $\langle \langle a_k, j \rangle, 1 \rangle$, for $1 \le j \le f_{i,k}$ [10]. In the sequel, the focus is on multisets, but the formalization and algorithms can be applied to sets and vectors.

Since this work focuses only on sets, multisets, and vectors, we only consider the similarity measures that exhibit the Shuffling Invariant Property (SIP). A measure exhibiting SIP is agnostic to the order of the elements in the alphabet $A$. Hence, shuffling the alphabet does not impact the similarity between multisets. For measures exhibiting SIP, the term Nominal Similarity Measures (NSMs) was coined in [8][4]. All the sets, multisets, and vectors similarity measures handled in the literature we are aware of are NSMs. For instance, the Jaccard similarity of two sets, $S_i$ and $S_j$, is given by $\frac{|S_i \cap S_j|}{|S_i \cup S_j|}$. The Ruzicka similarity [7] is the generalization of the Jaccard similarity to multisets. For any two multisets, $M_i \cap M_j = \sum_A \min(f_{i,k}, f_{j,k})$, and $M_i \cup M_j = \sum_A \max(f_{i,k}, f_{j,k})$. The set Dice similarity is given by $2 \times \frac{|S_i \cap S_j|}{|S_i|+|S_j|}$, and the set cosine similarity is given by $\frac{|S_i \cap S_j|}{\sqrt{|S_i| \times |S_j|}}$. Both Dice and cosine similarity can be trivially generalized to multisets using the set representation of multiset in [10]. The vector cosine similarity is given by $\frac{\sum_A |f_{i,k}| \times |f_{j,k}|}{|M_i| \times |M_j|}$. All these measures are agnostic to the order of the alphabet, and hence can be computed from partial results aggregated over the entire alphabet. More formally, NSMs can be expressed on the form of eqn. 1.

---
[3]Hadoop supports having different types for keys of the reducer input and output. The Google MapReduce does not.
[4]Similarity measures are surveyed in [7, 8, 15].

$$Sim(M_i, M_j) = F(\prod_{1}^{A}(g_1(f_{i,k}, f_{j,k})),$$
$$\prod_{2}^{A}(g_2(f_{i,k}, f_{j,k})),$$
$$\ldots$$
$$\prod_{L}^{A}(g_L(f_{i,k}, f_{j,k}))) \quad (1)$$

In eqn. 1, the $F()$ function combines the partial results of the $g_l(.,.)$ functions as aggregated over the alphabet by the $\prod_{l}^{A}$ aggregators, where $1 \le l \le L$, for some constant $L$.

### 3.2 Insight for High Scalability

The entire alphabet does not need to be scanned to compute the partial results combined using $F()$. We classify the $g_l(.,.)$ functions into three classes depending on which elements need to be scanned to compute the partial results.

The first *unilateral* class comprises functions whose partial results can be computed using a scan on the elements in only one multiset, either $U(M_i)$ or $U(M_j)$. Unilateral functions consistently disregard either $f_{i,k}$ or $f_{j,k}$. For instance, to compute the partial result $|M_i|$, $g_l(.,.)$ is set to the identity of the first operand, $f_{i,k}$, and $\prod_{l}^{A}$ to the $\sum$ aggregator. Scanning only the elements in $U(M_i)$, instead of the entire $A$, and applying the formula $\sum_{a_k \in U(M_i)} f_{i,k}$ yields $|M_i|$.

The second class of *conjunctive* functions can be computed using a scan on the elements in the intersection of the two multisets, $U(M_i \cap M_j)$. For instance, to compute the partial result $|M_i \times M_j|$, $g_l(.,.)$ is set to the multiplication function, and $\prod_{l}^{A}$ to the $\sum$ aggregator. Scanning only the elements in $U(M_i \cap M_j)$, instead of the entire $A$, and applying the formula $\sum_{a_k \in U(M_i \cap M_j)} f_{i,k} \times f_{j,k}$ yields $|M_i \times M_j|$.

Similarly, we define the class of *disjunctive* functions for those whose partial results can only be computed using a scan on the elements in the union of the two multisets, $U(M_i \cup M_j)$. For instance, to compute the symmetric difference, $|M_i \Delta M_j|$, $g_l(.,.)$ is set to the absolute of the difference, and $\prod_{l}^{A}$ to the $\sum$ aggregator. Scanning only the elements in $U(M_i \cup M_j)$, instead of the entire $A$, and applying the formula $\sum_{a_k \in U(M_i \cup M_j)} |f_{i,k} - f_{j,k}|$ yields $|M_i \Delta M_j|$.

Given this classification of functions, it is crucial to examine the complexity of accumulating the partial results of each of these classes. The partial results of the unilateral functions, denoted $Uni(M_i)$ for multiset $M_i$, can be accumulated for all multisets in a single scan on the dataset. The conjunctive partial results, denoted $Conj(M_i, M_j)$, can be accumulated for all pairs of multisets in a single scan on an

706

inverted index of the elements[5]. To compute the disjunctive partial results, for every pair of multisets that are candidates to be similar, their data needs to be scanned concurrently. Fortunately, all the similarity measures we are aware of can be expressed in terms of unilateral and conjunctive functions. We leave disjunctive functions for future work. All the published algorithms we are aware of, reviewed in § 6, cannot handle disjunctive function in the general case, since they generate candidate pairs from inverted indexes.

Some examples are given on expressing the widely used similarity measures in terms of unilateral and conjunctive functions. The Ruzicka similarity is given by $\frac{|M_i \cap M_j|}{|M_i \cup M_j|}$. Hence, the Ruzicka similarity is expressed in the form of eqn. 1 when $g_1(.,.)$ is the $\min(.,.)$ function, $g_2(.,.)$ is the $\max(.,.)$, both $\prod_{1}^{A}$ and $\prod_{2}^{A}$ are the $\sum$ aggregator, and $Sim(M_i, M_j)$ is $\frac{\sum_A g_1(f_{i,k}, f_{j,k})}{\sum_A g_2(f_{i,k}, f_{j,k})}$. Notice that the denominator contains the disjunctive function, $\max(.,.)$. Ruzicka can be rewritten as $\frac{|M_i \cap M_j|}{|M_i|+|M_j|-|M_i \cap M_j|}$, which is expressible in the form of eqn. 1 as $\frac{\sum_A g_1(f_{i,k}, f_{j,k})}{\sum_A g_2(f_{i,k}, f_{j,k}) + |g_3(f_{i,k}, f_{j,k})| - |g_1(f_{i,k}, f_{j,k})|}$, where $g_1(.,.)$ is the $\min(.,.)$ function, $g_2(.,.)$ and $g_3(.,.)$ are the identity of the first and second operand, respectively, and $\prod_{1}^{A}, \prod_{2}^{A}$, and $\prod_{3}^{A}$ are all $\sum$ aggregators. In this example, $Uni(M_i) = \langle |M_i| \rangle = \langle \sum_A g_2(f_{i,k}, f_{j,k}) \rangle$. Similarly, $Uni(M_j) = \langle |M_j| \rangle = \langle \sum_A g_3(f_{i,k}, f_{j,k}) \rangle$. Finally, $Conj(M_i, M_j) = \langle |M_i \cap M_j| \rangle = \langle \sum_A g_1(f_{i,k}, f_{j,k}) \rangle$. Similarly, the multiset cosine similarity, $\frac{|M_i \cap M_j|}{\sqrt{|M_i| \times |M_j|}}$, and the multiset Dice similarity, $2 \times \frac{|M_i \cap M_j|}{|M_i|+|M_j|}$, is expressed in the form of eqn. 1 by setting $g_1(.,.)$ to the $\min(.,.)$ function, $g_2(.,.)$ and $g_3(.,.)$ to the identity of the first and second operands, respectively, and setting the similarity function to $\frac{\sum_A g_1(f_{i,k}, f_{j,k})}{\sqrt{\sum_A g_2(f_{i,k}, f_{j,k}) \times \sum_A g_2(f_{i,k}, f_{j,k})}}$ for cosine, and $2 \times \frac{\sum_A g_1(f_{i,k}, f_{j,k})}{\sum_A g_2(f_{i,k}, f_{j,k}) \times \sum_A g_2(f_{i,k}, f_{j,k})}$ for Dice.

Given the above classification, in one pass over the dataset, the unilateral partial results, $Uni(M_i)$, can be accumulated for each $M_i$, and an inverted index can also be built. The inverted index can then be scanned to compute the conjunctive partial results, $Conj(M_i, M_j)$, for each candidate pair, $\langle M_i, M_j \rangle$, whose intersection is non-empty. The challenge is to join the unilateral partial results to the conjunctive partial results in order to compute the similarities.

## 4. THE V-SMART-JOIN FRAMEWORK

Instead of doing the join, the *V-SMART-Join* framework works around this scalability limitation. The general idea is to join $Uni(M_i)$ to all the elements in $U(M_i)$. Then, an inverted index is built on the elements in $A$, such that each entry of an element, $a_k$, has all the multisets containing $a_k$, augmented with their $Uni(.)$ partial results. For each pair of multisets sharing an element, $\langle M_i, M_j \rangle$, this inverted index contains $Uni(M_i)$ and $Uni(M_j)$. The inverted index can also be used to compute the $Conj(M_i, M_j)$. Hence, the inverted index can be used to compute $Sim(M_i, M_j)$ for all pairs.

The *V-SMART-Join* framework consist of two phases. The first *joining* phase joins $Uni(M_i)$ to all the elements in $U(M_i)$. The second *similarity* phase builds the inverted index, and computes the similarity between all candidate pairs. The algorithms of the joining phase are described in

---
[5]An inverted index groups all the multisets containing any specific element together.

§ 5. In this section, the focus is on the similarity phase, since it is shared by all the joining algorithms.

Each multiset, $M_i$, is represented in the dataset input to the *similarity* phase using multiple tuples, a tuple for each $a_k$, where $a_k \in M_i$. We call these input tuples on the form $\langle M_i, Uni(M_i), m_{i,k} \rangle$ *joined* tuples. This representation of the input data is purposeful. If each multiset is represented as one tuple, multisets with vast underlying cardinalities would cause scalability and load balancing problems.

The *V-SMART-Join* similarity phase is scalable, and comprises two MapReduce steps. The goal of the first step, $Similarity_1$, is to build the inverted index augmented with the $Uni(.)$ values, and scan the index to generate candidate pairs. The map stage transforms each entry of $m_{i,k}$ to be indexed by the element $a_k$, and caries down $Uni(M_i)$ and $f_{i,k}$ to the output tuple. The shuffler groups together all the tuples by their common elements. This implicitly builds an inverted index on the elements, such that the list of each element, $a_k$, is augmented with $Uni(M_i)$ and $f_{i,k}$ for each set $M_i$ containing $a_k$. For each element, $a_k$, a reducer receives a reduce_value_list$_{a_k}$. For each pair of multisets, $\langle M_i, M_j \rangle$ in reduce_value_list$_{a_k}$, the reducer outputs the identifiers, $\langle M_i, M_j \rangle$, along with $Uni(M_i)$, $Uni(M_j)$, $f_{i,k}$ and $f_{j,k}$. The map and reduce functions are formalized below.

map$_{Similarity_1}$:

$\langle M_i, Uni(M_i), m_{i,k} \rangle \to \langle a_k, \langle M_i, Uni(M_i), f_{i,k} \rangle \rangle$

reduce$_{Similarity_1}$:

$\langle a_k, (\langle M_i, Uni(M_i), f_{i,k} \rangle)^* \rangle \xrightarrow{\forall M_i, M_j \in \text{reduce\_value\_list}}$
$(\langle \langle M_i, M_j, Uni(M_i), Uni(M_j) \rangle, \langle f_{i,k}, f_{j,k} \rangle \rangle)^*$

The second step, $Similarity_2$, computes the similarity from the inverted index. It employs an identity map stage. A reducer receives reduce_value_list$_{\langle M_i, M_j \rangle}$ containing $\langle f_{i,k}, f_{j,k} \rangle$ for each common element, $a_k$ of a pair $\langle M_i, M_j \rangle$. The key of the list is augmented with $Uni(M_i)$ and $Uni(M_j)$. Therefore, $Similarity_2$ can compute $Conj(M_i, M_j)$, and combine it with $Uni(M_i)$ and $Uni(M_j)$ using $F()$. The result would be $Sim(M_i, M_j)$. Since computing the similarity of pairs of multisets with large intersections entails aggregation over long lists of $\langle f_{i,k}, f_{j,k} \rangle$ values, the lists are pre-aggregated using combiners to better balance the reducers' load. The map and reduce functions are formalized below.

map$_{Similarity_2}$:

$\langle \langle M_i, M_j, Uni(M_i), Uni(M_j) \rangle, \langle f_{i,k}, f_{j,k} \rangle \rangle \to$
 $\langle \langle M_i, M_j, Uni(M_i), Uni(M_j) \rangle, \langle f_{i,k}, f_{j,k} \rangle \rangle$

reduce$_{Similarity_2}$:

$\langle \langle M_i, M_j, Uni(M_i), Uni(M_j) \rangle, (\langle f_{i,k}, f_{j,k} \rangle)^* \rangle \to$
 $\langle M_i, M_j, Sim(M_i, M_j) \rangle$

Clearly, the performance of the similarity phase is little affected by changing the similarity measure, as long as the same $g_l(.,.)$ functions are used. That is, the impact of individual $g_l(.,.)$ functions onto the final similarity values does not affect the efficiency of the similarity phase.

The slowest $Similarity_1$ machine is the reducer that handles the longest reduce_value_list$_{a_k}$. The I/O time of this reducer is quadratic in $\max(Freq(a_k))$, the length of longest reduce_value_list$_{a_k}$. The longest reduce_value_list$_{a_k}$ also has to fit in memory to output the pairwise tuples, which may cause thrashing. The slowest $Similarity_2$ machine is the reducer that handles the longest intersection of all pairs of multisets. This $Similarity_2$ slowness is largely mitigated by using combiners, while the $Similarity_1$ slowness is not.

To speed up the slowest $Similarity_1$ reducer and avoid thrashing, elements whose frequency exceeds $q$, i.e., shared by more than $q$ multisets, for some relatively large $q$, can be discarded. These are commonly known as "stop words". Discarding stop words achieves better load balancing, is widely used in IR [5, 6, 13, 22, 29], and reduces the noise in the similarities when the elements have skewed frequencies, which is typical of Internet-traffic-scale applications. This can be done in a preprocessing MapReduce step. The preprocessing step maps input tuples from $\langle M_i, m_{i,k} \rangle$ to $\langle a_k, \langle M_i, f_{i,k} \rangle \rangle$. The preprocessing reducer buffers the first $q$ multisets in the reduce_value_list of $a_k$ and checks if the list was exhausted before outputting any $\langle M_i, m_{i,k} \rangle$ tuples. This way, the complexity of the slowest $Similarity_1$ reducer becomes quadratic in $q$ instead of $\max(Freq(a_k))$.

To avoid discarding stop words, avoid thrashing and still achieve high load balancing, the quadratic processing can be delegated from an overloaded $Similarity_1$ reducer to several $Similarity_2$ mappers. Each overloaded reducer can dissect its reduce_value_list into chunks of multisets, and output all possible pairs of chunks. Each pair of these chunks is read by a $Similarity_2$ mapper that would output all the possible pairs of the multisets in this pair of chunks.

To achieve this, the reducers have to make use of the capability of rewinding their reduce_value_lists. A $Similarity_1$ reducer that receives an extremely long reduce_value_list can dissect this list into $T$ large chunks, such that each chunk consumes less than $\frac{B}{2}$ Bytes, where $B$ is the available memory per machine, for some $T$. Each chunk is on the form $\langle a_k, (\langle M_i, Uni(M_i), f_{i,k} \rangle)^* \rangle$. The reducer outputs all the possible $T^2$ pairs of chunks in a nested loop manner, which entails rewinding the input $T$ times. The output of such a reducer will be different from the other normal $Similarity_1$ reducers, and can be signaled using a special flag.

These $T^2$ pairs of chunks can fit in memory and can be processed by up to $T^2$ different $Similarity_2$ mappers. Instead of acting as identity mappers, the $Similarity_2$ mappers process their input in a way similar to the normal $Similarity_1$ reducers when receiving pairs of chunks, $\langle Chunk_p, Chunk_q \rangle$, where $1 \leq p, q \leq T$. That is, when the input is on the form $\langle \langle a_k, (\langle M_i, Uni(M_i), f_{i,k} \rangle)^* \rangle, \langle a_k, (\langle M_j, Uni(M_j), f_{j,k} \rangle)^* \rangle \rangle$, it outputs $\langle \langle M_i, M_j, Uni(M_i), Uni(M_j) \rangle, \langle f_{i,k}, f_{j,k} \rangle \rangle$ for each $M_i \in Chunk_p$, and each $M_j \in Chunk_q$. This better balances the load among the $Similarity_1$ reducers while not skewing the load among the $Similarity_2$ mappers, without discarding stop words. In addition, the I/O cost of the slowest $Similarity_1$ reducer becomes proportional to $T \times \max(Freq(a_k))$ instead of $\max(Freq(a_k))^2$.

## 5. THE JOINING PHASE ALGORITHMS

This section describes the joining algorithms that, for each $M_i$, join $Uni(M_i)$ to its elements. In other words, it transforms the *raw* input tuples on the form $\langle M_i, m_{i,k} \rangle$ to joined tuples on the form $\langle M_i, Uni(M_i), m_{i,k} \rangle$.

### 5.1 The Online-Aggregation Algorithm

For each input tuple, the mapper outputs the information necessary to compute $Uni(M_i)$ with secondary key 0, as well as the same exact input tuple with secondary key 1. For each multiset $M_i$, a reducer receives reduce_value_list$_{M_i}$ with the output of the mappers sorted by the secondary key. The reducer scans reduce_value_list$_{M_i}$, and computes $Uni(M_i)$, since the information for this computation, secondary keyed by 0, comes first in reduce_value_list$_{M_i}$. The reducer then continues to scan the elements, secondary keyed by 1, and outputs the multiset id, $M_i$ with the computed partial result, $Uni(M_i)$, with each element $m_{i,k}$. The map and reduce functions are formalized below.

map$_{Online-Aggregation_1}$:

$$\langle M_i, m_{i,k} \rangle \xrightarrow{\text{if } f_{i,k} > 0} \langle M_i, 0, f_{i,k} \rangle, \langle M_i, 1, m_{i,k} \rangle$$

reduce$_{Online-Aggregation_1}$:

$$\langle M_i, (0, (f_{i,k})^*), (1, (m_{i,k})^*) \rangle \to (\langle M_i, Uni(M_i), m_{i,k} \rangle)^*$$

The *Online-Aggregation* is very scalable, straightforward, and achieves excellent load balancing due to using combiners. However, it assumes the shuffler sorts the reducer input by the secondary keys for sorting. As discussed in § 2, Hadoop provides no support for secondary keys, and the workarounds are either unscalable, or entails writing parts of the engine. Even more, we could not find any published instructions on how to use the combiners with the secondary keys workarounds in a scalable way. Next, we propose other scalable algorithms that can be executed on Hadoop, and compare the performance of all the algorithms in § 7.

### 5.2 The Lookup Algorithm

The *Lookup* algorithm consists of two steps. The first $Lookup_1$ step computes $Uni(M_i)$ for each $M_i$. The mapper outputs $f_{i,k}$ keyed by $M_i$ for each input tuple $M_i, m_{i,k}$. The reducers scan a reduce_value_list$_{M_i}$, and compute $Uni(M_i)$ for each $M_i$. The output of the reducers are files mapping each $M_i$ to its $Uni(M_i)$. Combiners are also used here to improve the load balancing among reducers. The map and reduce functions are formalized below.

map$_{Lookup_1}$:

$$\langle M_i, m_{i,k} \rangle \xrightarrow{\text{if } f_{i,k} > 0} \langle M_i, f_{i,k} \rangle$$

reduce$_{Lookup_1}$:

$$\langle M_i, (f_{i,k})^* \rangle \to \langle M_i, \langle Uni(M_i) \rangle \rangle$$

When a mapper of the second step, $Lookup_2$, starts, it loads the files produced by $Lookup_1$ into a memory-resident lookup hash table. As each $Lookup_2$ mapper scans an input tuple, $\langle M_i, m_{i,k} \rangle$, it joins it to $Uni(M_i)$ using the lookup table. The output of the mappers of $Lookup_2$ is the same as the output of the mappers of $Similarity_1$. Hence, the $Similarity_1$ reducer can process the files output by the $Lookup_2$ mappers directly. The map function is formalized below.

map$_{Lookup_2}$:

$$\langle M_i, m_{i,k} \rangle \xrightarrow{lookup} \langle a_k, \langle M_i, Uni(M_i), f_{i,k} \rangle \rangle$$

The *Lookup* algorithm suffers from limited scalability. The second step assumes that the results of the first step can be loaded in memory to be used for lookups. If the memory cannot accommodate a lookup table with an entry for each $M_i$, the reducers suffer from thrashing. We next propose the *Sharding* algorithm that avoids this scalability limitation.

## 5.3 The Sharding Algorithm

The *Sharding* algorithm is a hybrid one between *Online-Aggregation* and *Lookup*. It exploits the skew in the underlying cardinalities of the multisets to separate the multisets into *sharded* and *unsharded* multisets. Sharded multisets have vast underlying cardinalities, are few in numbers, and are handled by multiple machines in a manner similar to *Lookup* without sacrificing scalability. Any unsharded multiset can fit in memory, and is handled in a way similar to the *Online-Aggregation* algorithm.

The *Sharding* algorithm consists of two steps. The first $Sharding_1$ step is the same as $Lookup_1$, with one exception. The reducer computes $Uni(M_i)$, and outputs a mapping from $M_i$ to its $Uni(M_i)$ only for each multisets, $M_i$, whose $|U(M_i)| > C$, for some parameter $C$. The map and reduce functions are formalized below.

$\text{map}_{Sharding_1}$:

$$\langle M_i, m_{i,k} \rangle \xrightarrow{\text{if } f_{i,k} > 0} \langle M_i, f_{i,k} \rangle$$

$\text{reduce}_{Sharding_1}$:

$$\langle M_i, (f_{i,k})^* \rangle \xrightarrow{\text{if } |U(M_i)| > C} \langle M_i, \langle Uni(M_i) \rangle \rangle$$

At the beginning of $Sharding_2$, each mapper loads the output of the $Sharding_1$ step to be used as a lookup table, exactly like the case of $Lookup_2$. As each $Sharding_2$ mapper scans an input tuple, $\langle M_i, m_{i,k} \rangle$, it joins it to $Uni(M_i)$ using the lookup table. If the join succeeds, it is established that $|U(M_i)| > C$, and $M_i$ is a sharded multiset. The mapper computes the fingerprint of $a_k$, and outputs the joined tuple keyed by $\langle M_i, fingerprint(a_k) \rangle$. The goal of adding $fingerprint(a_k)$ to the index is to distribute the load randomly among all the reducers. If the join fails, it is established that $|U(M_i)| \leq C$, and hence, a list of all the elements in $U(M_i)$ can fit in memory. In that case, the joined tuple keyed by $\langle M_i, -1 \rangle$ is output. Since the second entry in the tuple is always $-1$, all the elements from $M_i$ will be consumed by the same $Sharding_2$ reducer. Since reduce_value_list$_{M_i}$ fits in memory, the reducers can compute $Uni(M_i)$, and join it to the individual elements in $U(M_i)$.

A $Sharding_2$ reducer receives either a tuple with $Uni(M_i)$ joined in if $M_i$ is sharded, or a tuple with no joined $Uni(M_i)$ if $M_i$ is unsharded. If the tuple has the $Uni(M_i)$ information, the reducer strips off the fingerprint, and outputs a joined tuple for each element. If the tuple does not contain $Uni(M_i)$, then $M_i$ is unsharded, and reduce_value_list$_{M_i}$ fits in memory. The reducer loads reduce_value_list$_{M_i}$ in memory and scans it twice. The first time to compute $Uni(M_i)$, and the second time to output a joined tuple on the form $\langle M_i, Uni(M_i), m_{i,k} \rangle$ for each element $a_k$ in $U(M_i)$. The map and reduce functions are formalized below.

$\text{map}_{Sharding_2}$:

$$\langle M_i, m_{i,k} \rangle \xrightarrow{\text{if } f_{i,k} > 0}$$
$$\begin{cases} \xrightarrow{lookup} & \langle \langle M_i, fingerprint(a_k) \rangle, \langle Uni(M_i), m_{i,k} \rangle \rangle \\ & \text{if } M_i \in \text{Lookup} \\ \xrightarrow{lookup} & \langle \langle M_i, -1 \rangle, \langle NULL, m_{i,k} \rangle \rangle \\ & \text{if } M_i \notin \text{Lookup} \end{cases}$$

$\text{reduce}_{Sharding_2}$:

$$\langle \langle M_i, fingerprint(a_k) \rangle, (\langle Uni(M_i), m_{i,k} \rangle)^* \rangle \to$$
$$\langle M_i, Uni(M_i), m_{i,k} \rangle$$
$$\langle \langle M_i, -1 \rangle, (\langle NULL, m_{i,k} \rangle)^* \rangle \to \langle M_i, Uni(M_i), m_{i,k} \rangle$$

The *Sharding* algorithm is scalable, and is largely insensitive to the parameter $C$, as shown in § 7. The main goal of the parameter $C$ is to separate the very few multisets with vast underlying cardinalities that cannot fit in memory from the rest of the multisets. This separation of multiset is critical for the scalability of the algorithm. Therefore, the use of $C$ should not be nullified by setting $C$ to trivially large or small values. Setting $C$ to a huge value stops this separation of multisets into sharded and unsharded categories. In that case, $Sharding_1$ reducers processing multisets with vast underlying cardinalities would be overly loaded, and would suffer from thrashing. Conversely, setting $C$ to a trivially small value transforms the algorithm into a lookup algorithm, and the $Sharding_2$ mappers will have to fit in memory a lookup table mapping almost each $M_i$ to its $Uni(M_i)$.

For the three proposed algorithms, the slowest machine is the reducer that handles the multiset with the largest underlying cardinality. The I/O cost of these reducers is proportional to $max(|U(M_i)|)$. However, this slowness is greatly reduced by using combiners. Dedicated combiners are used in every aggregation to conserve the network bandwidth.

It is also worth noting that for any two measures that use the same $g_l(.,.)$ functions (e.g., Dice and cosine), the performance of the joining algorithms is little affected by using one over the other.

Next, the related work is discussed with a special focus on the *VCL* algorithm [33]. *VCL* is used as a baseline to evaluate the performance and scalability of the proposed algorithms in § 7.

## 6. RELATED WORK

Related problems have been tackled in different applications, programming paradigms, and using various similarity measures for sets, multisets, and vectors. This section starts by a general review, and then discusses *VCL* in details.

### 6.1 All-Pair Similarity Join Algorithms

Several approximate sequential algorithms employ Locality Sensitive Hashing (LSH), whose key idea is to hash the elements of the sets so that collisions are proportional to their similarity [18]. An inverted index is built on the union of hashed elements in all the sets. The goal is to avoid the quadratic step of calculating the similarity between all sets unless it is absolutely necessary.

Broder *et al.* proposed a sequential algorithm to estimate the Jaccard similarity between pairs of documents [5, 6] using LSH. In [5, 6], each document is represented using a set, $S_i$, comprising all its shingles, where a shingle is a fixed-length sequence of words in the document. A more scalable version of the algorithm is given in [22] in the context of detecting attacks from colluding attackers. The LSH process was repeated using several independent hash functions to establish probabilistic bounds on the errors in the similarity estimates. While these algorithms considered sets only, they can employ the set representation of multiset proposed in [10] to estimate the generalized Ruzicka similarity.

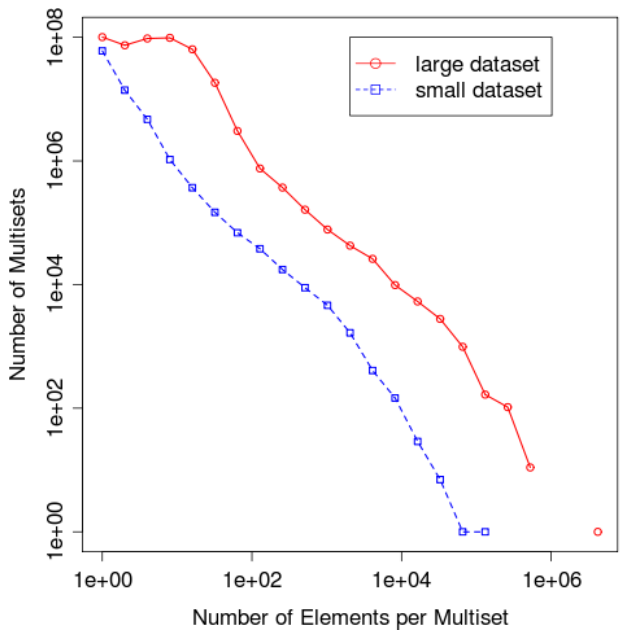

Figure 2: The distribution of elements per multiset.

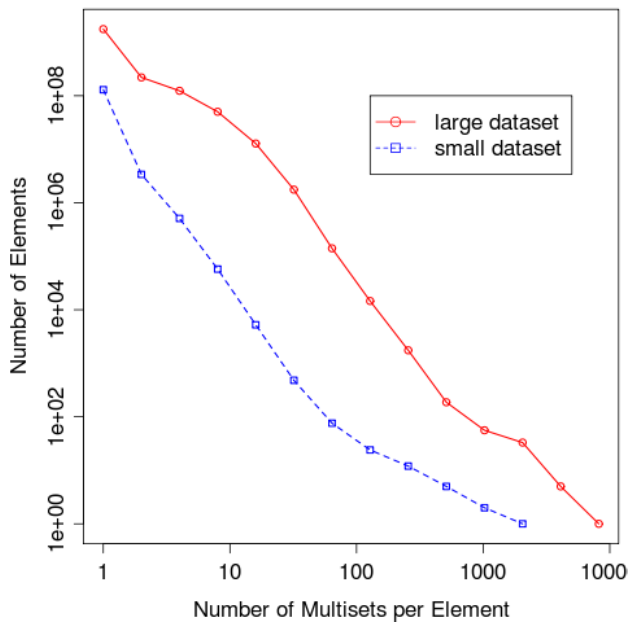

Figure 3: The distribution of multisets per element.

LSH was also used in [9] to approximate other similarity measures such as the Earth Mover Distance (EMD) between distributions[6] [28], and the cosine similarity between sets. However, the estimated similarities have a multiplicative bias that grows linearly with $\log(|A|)\log\log(|A|)$, which might be impractical for large alphabets, such as cookies[7].

Using inverted indexes is proposed to solve the all-pair similarity join problem exactly in [29]. Instead of scanning the inverted index and generating all pairs of sets sharing an element, the algorithm in [29] proceeds in two phases. The first candidate generation phase scans the data, and for each set, $S_i$, selects the inverted index entries that correspond to its elements. The algorithm then sorts the elements in this partial index by their frequency in order to exploit the skew in the frequencies of the elements. The algorithm dissects these elements into two partial indexes. The first partial index comprises the least frequent elements (i.e., elements with short lists of sets), and is denoted $Prefix(S_i)$. The second index comprises the most frequent elements (i.e., elements with long lists of sets), and is denoted $Suffix(S_i)$. The length of $Suffix(S_i)$ is determined based on $|S_i|$ and $t$, such that the similarity between $S_i$ and any other set cannot be established using only all the elements in the suffix. The candidate generation phase merges all the lists in the prefix and generates all the candidates that may be similar to $S_i$. In the second verification phase, the candidates are verified using the elements in the suffix. By dissecting the partial index of $S_i$ into a prefix and a suffix, the threshold $t$ is exploited and the expensive step of generating all the candidates sharing any element in their suffixes is avoided.

Several pruning techniques were proposed to further reduce the number of candidates generated. One such prominent technique is *prefix filtering* [10, 4, 34]. The technique builds an inverted index only for the union of the prefix elements of all the sets, which reduces the size of the inverted indices by a approximately $1-t$, according to [34]. Similarly, [34] proposed suffix filtering. In fact, [34] bundled prefix filtering and suffix filtering into a state of the art sequential algorithm, PPJoin+, along with positional filtering (the positions of the elements in any pair of overlapping ordered sets can be used to upper bound their similarity), and size filtering [2] (similar sets have similar sizes from the pigeonhole concept). Integrating most of these pruning techniques algorithmically was investigated in [19].

The MapReduce-based algorithm in [13] approximate the multiset similarity using the vector cosine similarity. The algorithm and the approximation is adopted in [3] with optimizations borrowed from [4] to reduce the communication between the machines and distribute the load more evenly. These techniques represent multisets as unit vectors, which ignores their cardinalities. This approximation allows for devising simple MapReduce algorithms. However, these techniques are not applicable when multisets are skewed in size, and the sizes of the multisets are relevant, which is typical in Internet-traffic application. In addition, these techniques provide approximate similarities, which obviates the use of the MapReduce framework that can be used to crunch large datasets to provide exact results.

The PPJoin+ algorithm is adopted in a MapReduce setting in [33] for database joins. Since this is the only algorithm that is exact, distributed, and versatile, it is used as a benchmark and is explained in details next.

---

[6]Given two piles of dirt in the shapes of the distributions, the distance measure is proportional to the effort to transform one pile into the other.

[7][16] has reported the bias factor grows linearly with $|A|$. In another analysis [17], Henzinger reported that the algorithm in [9] is more accurate than the algorithm in [5, 6] on the application of detecting near-duplicate web pages when using the same fingerprint size. That is attributed to the ability of [9] to respect the repeated shingles in the documents. The number of independent hash functions used in [17] is 84. It is notable that this is significantly less than the number of hash functions proposed in [22] of 423 to guarantee an error bound of 4% with confidence 95%. Clearly, [17] did not consider the set representation of multisets described in [10].

## 6.2 The VCL Algorithm

The *VCL* algorithm[8] was devised for set similarity joins where the sets come from two different sources. The algorithm was also adapted to solve the all-pair similarity join problem where the sets come from the same source, which is the problem in hand. While the work in [33] targets sets, it is applicable to multisets and vectors.

*VCL* is a MapReduce adaptation of PPJoin+ proposed in [34] that reduces the number of candidate pairs by combining several optimizations. In fact, the main MapReduce step of *VCL* relies on prefix filtering, explained in § 6.1. To apply the candidate pairs filtering technique [34], *VCL* makes a preprocessing scan on the dataset to sort the elements of the alphabet, $A$, by frequency. During the initialization of the mappers of the main phase, all the elements, sorted by their frequencies, are loaded into the memory of the mappers.

Each mapper processes a multiset at a time, and each multiset is processed by one mapper. For each multiset, $M_i$, the mapper computes the prefix elements of $M_i$, and outputs the entire content of $M_i$ with each element $a_k \in \textit{Prefix}(M_i)$. *VCL* uses the MapReduce shuffle stage to group together multisets that share any prefix element. Hence, each reducer receives a reduce key, element $a_k$, along with the reduce_value_list$_{a_k}$ comprising all the multisets for which $a_k$ is a prefix element. For each multiset in the reduce_value_list$_{a_k}$, the reducer has the elements of the entire multiset, and can compute the similarity between each pair of multisets. This algorithm computes the similarity of any two multisets on each reducer processing any of their common prefix elements. These similarities are deduplicated in a post-processing phase. The map and reduce functions of the kernel, i.e., main, phase are formalized below.

map$_{VCL}$:

$$\langle M_i, \{m_{i,1}, \ldots, m_{i,|A|}\}\rangle \xrightarrow{\forall a_k \in \textit{Prefix}(M_i)}$$
$$(\langle a_k, \langle M_i, \{m_{i,1}, \ldots, m_{i,|A|}\}\rangle\rangle)^*$$

reduce$_{VCL}$:

$$\langle a_k, (\langle M_i, \{m_{i,1}, \ldots, m_{i,|A|}\}\rangle)^*\rangle \xrightarrow{\forall M_i, M_j \in \text{ reduce\_value\_list}}$$
$$(\langle M_i, M_j, \textit{Sim}(M_i, M_j)\rangle)^*$$

*VCL* suffers from major inefficiencies in the computation, network bandwidth, and storage. For each multiset, $M_i$, the map stage incurs a network bandwidth and storage cost that is proportional to $|\textit{Prefix}(M_i)| \times |U(M_i)|$. Hence, the map bottleneck is the mapper handling the largest multiset. This constituted a major bottleneck in the reported experiments. In addition, the reducers suffer from high redundancy. Each pair of multisets, $M_i$ and $M_j$, have their similarity computed $|\textit{Prefix}(M_i) \cap \textit{Prefix}(M_j)|$ times. This inefficiency cannot be alleviated using combiners.

To reduce this inefficiency, grouping of elements into super-elements was proposed in [33]. Representing multisets in terms of super-elements shrinks the multisets, and hence reduces the network, memory, and disk footprint. Grouping elements shrinks the alphabet, and hence a list of the super-elements, sorted by their frequencies, can be more easily accommodated in the memories of the *VCL* kernel mappers. In addition, grouping reduces the number of kernel reducers calculating the similarity of pairs of multisets. The kernel reducers produce a candidate pair of multisets if their similarity of super-elements exceeds the threshold, $t$. Grouping produces "superfluous" pair of multisets that can share a prefix super-element, while not sharing a prefix element. These superfluous pairs are weeded out in the post-processing phase. In the experiments in [33], grouping was shown to consistently introduce more overhead than savings due to the superfluous pairs, and the authors suggested using one element per group. This renders the *VCL* algorithm incapable of handling applications where the alphabet has to fit completely in memory of the mappers.

The *VCL* algorithm suffers from another major scalability bottleneck. In the map function of the kernel phase and the post-processing phase, entire multisets are read, processed, and output as whole indivisible capsules of data. Hence, *VCL* can only handle multisets that can fit in memory. This renders the algorithm inapplicable of handling Internet-traffic-scale applications, where the alphabet could be the cookies visiting Google, and the multisets could be the IPs visiting Google with these cookies.

## 7. EXPERIMENTAL RESULTS

To establish the scalability and efficiency of the *V-SMART-Join* algorithms, experiments were carried out with datasets of real IPs and cookies. Each IP was represented as a multiset of cookies, where the multiplicity is the number of times the cookie appeared with an IP. The similarity measure used was Ruzicka. The experiments were conducted using two datasets from the search query logs. The first dataset is of much smaller size and it had approximately 133 Million unique elements (cookies) shared by approximately 82 Million multisets (IPs). The first dataset was used so that all the algorithms can finish processing it. This smaller dataset was used as a litmus test to know which algorithms will be compared on the second dataset.

The second dataset is of a more realistic size, and is used to know which algorithms can solve the all-pair similarity join problem in an Internet-traffic-scale setting, and compare their efficiency. The second dataset had approximately 2.2 Billion unique elements (cookies) shared by approximately 454 Million multisets (IPs). The distributions of the multisets and elements are plotted in Fig. 2 and Fig. 3.

Clearly, both the multisets, the IPs, and the alphabet, the cookies, are in the order of hundreds of millions to billions. In addition, the distributions are fairly skewed. However, no stop words were discarded, and no multisets were sampled.

The algorithms analyzed in this experimental evaluation are the proposed algorithms as well as the state of the art algorithm, *VCL*. We did not include the LSH-based algorithms since the existing algorithms are serial, and generalizing them to a distributed setting is beyond the scope of this work. In addition, LSH algorithms are approximate. Using the computing power of multiple machines in a parallel setting obviates the need to approximation, especially if the exact algorithms can finish within reasonable time.

All the algorithms were allowed 1GB of memory, and 10GB of disk space on each of the machines they ran on, and they all ran on the same number of machines. All the algorithms were started concurrently to factor out any measurement biases caused by the data center loads. All the reported run times represent a median-of-5 measurements.

---

[8]The algorithm is referred to as *VCL* after the names of the authors of [33].

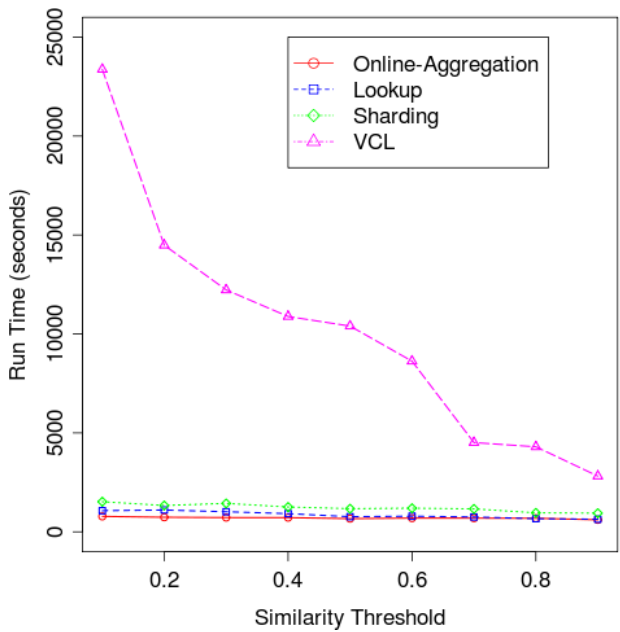
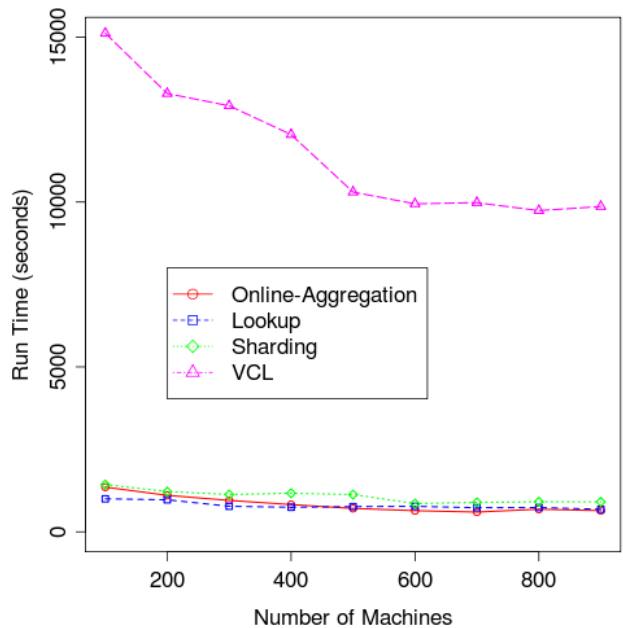

Figure 4: Algorithms run time on the small dataset with various similarity thresholds (500 machines).

Figure 5: Algorithms run time on the small dataset with various numbers of machines ($t = 0.5$).

The results of comparing the algorithms on the small and realistic datasets are reported in § 7.1 and § 7.2, respectively. We also conduct a sensitivity analysis of the *Sharding* algorithm with respect to the parameter $C$ in § 7.3. Finally, we briefly comment on discovering load balancers in § 7.4.

## 7.1 Algorithms Comparison on the Small Dataset

The first step in comparing the algorithms on the small dataset was to run each algorithm on the same number of machines, 500, and to vary the similarity threshold, $t$, between 0.1 and 0.9 at an 0.1 interval. Understandably, all the algorithms produced the same number of similar pairs of IPs for each value of $t$. The results are plotted in Fig. 4. Clearly, the performance of the *VCL* algorithm in terms of run time was not close to any of the *V-SMART-Join* algorithms. In addition, its performance was highly dependent on the similarity threshold, $t$. It is also worth mentioning that at least 86% of the run time of *VCL* was consumed by the map phase of the kernel MapReduce step, where the multisets get replicated for each prefix element. The *V-SMART-Join* algorithms were fairly insensitive to $t$. Their run time decreased very slightly as $t$ increased, since less pairs were output, which reduces the I/O time.

The *Online-Aggregation* algorithm was consistently the most efficient. *Online-Aggregation* executed 30 times faster than *VCL* when the similarity threshold was 0.1. When the threshold was increased to 0.9, the performance of *VCL* improved to be only 5 times worse than *Online-Aggregation*. *Online-Aggregation* was followed by *Lookup*, and then *Sharding*, with slight differences in performance. This was expected, since the *Online-Aggregation* joining needs only one MapReduce step. The *Lookup* algorithm saves a MapReduce step compared to the *Sharding* algorithm.

How the algorithms scale out relative to the number of machines was also examined. All the algorithms were run to find all pairs of similarity 0.5 or more, and the number of machines were varied from 100 to 900 at an interval of 100 machines. Again, the *VCL* algorithm performed a lot worse than the *V-SMART-Join* algorithms. In addition, when the algorithm ran on over 500 machines, it did not make much use of the machines. The reason is that the bottleneck of the runs was outputting each large multiset with each one of its prefix elements. This results in a huge load unbalance. That is, some of the machines that handle the large multisets become very slow, which is independent of the number of machines used. When using 900 machines instead of 100 machines, *VCL* run time dropped by 35%.

On the other hand, the *V-SMART-Join* algorithms continued to observe a relative reduction in the run time as more machines were used. This speed up was hampered by the fact that a large portion of the run times were spent in starting and stopping the MapReduce runs. The algorithm that exhibited the most reduction in run time was *Online-Aggregation*, whose run time dropped by 53%, while the *Lookup* showed the least reduction in run time with a drop of 32%. This is because part of the run time of *Lookup* was loading the lookup table mapping each $M_i$ to $Uni(M_i)$ on each machine, which is a fixed overhead regardless of the number of machines used. Again, *Online-Aggregation* outperformed *VCL* by 11 to 15 times depending on the similarity threshold.

## 7.2 Algorithms Comparison on the Realistic Dataset

The algorithms were run on the more realistic dataset, and the results are presented below. It is worth mentioning that *Lookup* did not succeed because it was never able to load the entire lookup table mapping each $M_i$ to $Uni(M_i)$. Hence, *Lookup* was out of the competition. Similarly, the *VCL* algorithm was not able to load all the cookies, sorted by their frequency. To remedy this, the cookie elements were sorted based on their hash signature instead of their frequencies. However, even with this modification, *VCL* never finished the runs within two days. The mappers of the kernel step took more than 48 hours to finish, and were killed by the MapReduce scheduler.

The remaining algorithms, *Online-Aggregation* and *Sharding*, were compared. The similarity phase is common to both algorithms. Hence, the time for running the joining phase was measured separately from the time for running the similarity phase. Since these algorithms do not get affected by the similarity threshold, only their scaling out with the number of machines was compared. The algorithms were run to find all pairs of similarity 0.5 or more, and the number of machines were varied from 100 to 900 at an interval of 100 machines. The results are plotted in Fig. 6. From the figure, both algorithms, as well as the common similarity step were able to scale out as the number of machines increased. *Online-Aggregation* took roughly half the time of *Sharding*.

### 7.3 How Sensitive is Sharding to $C$?

The previous section shows that while the *Sharding* algorithm is half as efficient as the *Online-Aggregation* algorithm, it is still scalable. The main advantage of *Sharding* is it does not use secondary keys, which are not supported natively by Hadoop. On the other hand, *Sharding* takes a parameter $C$. The function of parameter $C$ is to separate the multisets with vast underlying cardinalities, whose $Uni(.)$ functions are calculated and loaded in memory as the $Sharding_2$ mappers start, from the rest of the multisets, whose $Uni(.)$ are calculated on the fly by the $Sharding_2$ reducers. A sensitivity analysis was conducted on the performance of the *Sharding* algorithm as the parameter $C$ was varied. The run time of the $Sharding_1$ and $Sharding_2$ steps, as well as their sum, are plotted in Fig. 7 as the parameter $C$ is varied between $2^5$ and $2^{15}$ using exponential steps.

The run time of the $Sharding_1$ step decreased since less pairs were output as $C$ increased, which reduced the I/O time. On the other hand, the run time of the $Sharding_2$ step increased since more on the fly aggregation is done as $C$ increased. The total run time of the *Sharding* algorithm stayed stable throughout entire range of $C$. More precisely, the total run time had a slight downward trend until the value of $C$ was roughly 1000 and then increased again. Notice however that larger values of $C$ reduce the memory footprint of the algorithm, and are then more recommended.

### 7.4 A Comment on Identifying Proxies

We conclude the experimental section by briefly discussing the discovered IP communities. For each similarity threshold, a manual analysis was done on a random sample of the similar IPs. Each threshold was judged based on its coverage, i.e., the number of discovered similar IPs, and the false positives of the sample. False positives are defined as IPs in the results that cannot be proxies. Similar IPs are judged as not proxies based on evidences independent of this study. An example is the case when two IPs that were judged by this approach to be similar belong in fact to two different organizations. Clearly, setting $t$ to 0.1 yields the highest coverage, but also the highest false positives.

To reduce the false positives, instead of reducing the similarity threshold, IPs that observed less than 50 cookies were filtered out. This almost eliminated the false positives for all the thresholds, since it eliminated all the IPs that have very low chance of acting as proxies. After eliminating these IPs, the number of cookies were around two orders of magnitude larger than the number of IPs. It is expected to find a lot more cookies than IPs in proxy settings.

Notice that this filtering of small IPs would not improve the reported performance of *VCL*, though it would improve the reported performance of *Lookup*. The reason is the main bottleneck of *VCL* are multisets with vast underlying cardinalities. These bottleneck multisets are the most important to identify in order to discover load balancer, and should not be filtered out. On the other hand, by reducing the number of multisets, the *Lookup* algorithm reduces the I/O time of $reduce_{Lookup_1}$ responsible for producing the data for the lookup table mapping each $M_i$ to $Uni(M_i)$. It is also worth noting that this filtering allowed the *Lookup* algorithm to accommodate the lookup table of the realistic dataset, and was able to finish the run in time very comparable to the *Online-Aggregation* algorithm.

The overwhelming majority of the discovered load balancers were in European countries. The seven largest strongly connected sets of IPs spanned several subnetworks, and comprised thousands of IPs. The load balancers in Saudi Arabia and North Korea were few, but were the most active.

## 8. DISCUSSION

The *V-SMART-Join* MapReduce-based framework for discovering all pairs of similar entities is proposed. This work presents a classification of the partial results necessary for calculating Nominal Similarity Measures (NSMs) that are typically used with sets, multisets, and vectors. This classification enables splitting the *V-SMART-Join* algorithms into two stages. The first stage computes and joins the partial results, and the second stage computes the similarity for all candidate pairs. The *V-SMART-Join* algorithms were up to 30 times as efficient as the state of the art algorithm, *VCL*, when compared on real small datasets. We also established the scalability of the *V-SMART-Join* algorithms by running them on a dataset of a realistic size, on which the *VCL* mapper never succeeded to finish, not even when *VCL* was modified to improve scalability.

We touch on the reason why we did not incorporate prefix filtering into the proposed algorithms. While prefix filtering reduces the generated candidates from any pair of multisets sharing an element to only those that share a prefix element, employing it in a MapReduce algorithm introduces a scalability bottleneck, which defeats the purpose of using MapReduce. First, loading a list of all the alphabet elements, sorted by their frequencies, in memory to identify the prefix elements of each entity renders prefix filtering inappropriate for handling extremely large alphabets. This was a bottleneck for the algorithms in [3, 33]. Extremely large alphabets and entities are common in Internet-traffic-scale applications. While [33] proposed grouping elements to reduce the memory footprint of prefix filtering, their experiments showed the inefficiencies introduced by grouping. Second, the approach of generating candidates and then verifying them entails machines loading complete multisets as indivisible capsules. This limits the algorithms in [3, 33] to datasets where pairs of multisets can fit in memory. Finally, as clear from the experiments, prefix filtering is only effective when the similarity threshold is extremely high. Prefix filtering becomes less effective when the similarity threshold drops. As was clear from our application, the threshold was set to a small value (0.1) to find all similar IPs, which minimizes the benefits of prefix filtering.

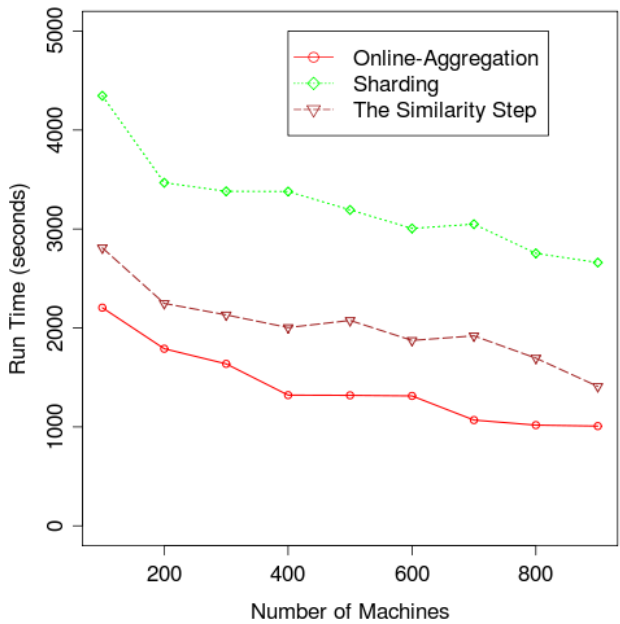
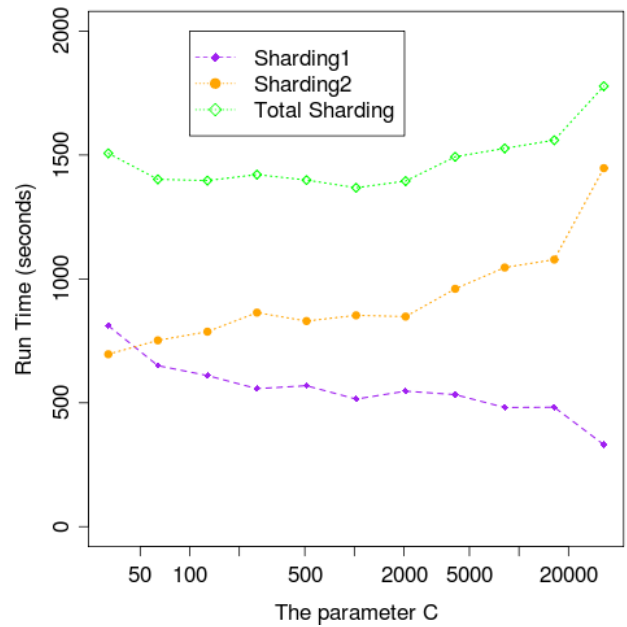

Figure 6: Algorithms run time on the large dataset with various numbers of machines ($t = 0.5$).

Figure 7: The run time of *Sharding* on the large dataset with various values of the parameter $C$.

The main lesson learned from this work is that devising new algorithms for the MapReduce setting may yield algorithms that are more efficient and scalable than those devised by adopting sequential algorithms for this distributed setting. Adopting sequential algorithms to the distributed settings may overlook capabilities and functionalities offered by the MapReduce framework. It is also crucial to devise algorithms that are compatible with the publicly available version of MapReduce, Hadoop, for wider adoption.

Finally, it is constructive to identify the limitations of this work. The proposed algorithms, as well as others in the literature, handles only NSMs whose partial results can be computed either by scanning the two entities, or by scanning the intersection of the two entities. That is, the algorithms do not handle NSMs if any of its partial results entail scanning the elements in the union of the two entities. This still makes this work applicable to a large array of similarity measures, such as Jaccard, Ruzicka, Dice, and cosine.

In addition, this work assumes large scale datasets with numerous entities, numerous elements, and a skew in the sizes of the entities. The skew in the sizes of the entities enabled the sharding algorithm to categorize entities into sharded and unsharded entities. This work is not applicable to datasets with numerous entities and very few elements. For instance, if the entities represent distribution histograms of a moderate number bins, and the elements represent the bins, almost each bin would be shared by almost all the entities. In that case, the algorithm would have to do an exhaustive pairwise similarity join, which is very unscalable. Our future work focuses on devising a MapReduce-based algorithm for all-pair similarity joins of histograms.

## Acknowledgments

We would like to thank Matt Paduano for his valuable discussions, Amr Ebaid for implementing the *Lookup* algorithm, Adrian Isles and the anonymous reviewers for the rigorous revision of the manuscript.

## 9. REFERENCES

[1] Apache Hadoop. http://hadoop.apache.org.
[2] A. Arasu, V. Ganti, and R. Kaushik. Efficient Exact Set-Similarity Joins. In *Proceedings of the 32nd VLDB International Conference on Very Large Data Bases*, pages 918–929, 2006.
[3] R. Baraglia, G. De Francisci Morales, and C. Lucchese. Document Similarity Self-Join with MapReduce. In *Proceedings of the 10th IEEE ICDM International Conference on Data Mining*, pages 731–736, 2010.
[4] R. Bayardo, Y. Ma, and R. Srikant. Scaling Up All Pairs Similarity Search. In *Proceedings of the 16th WWW International Conference on World Wide Web*, pages 131–140, 2007.
[5] A. Broder. On the Resemblance and Containment of Documents. In *Proceedings of the IEEE SEQUENCES Compression and Complexity of Sequences*, pages 21–29, 1997.
[6] A. Broder, S. Glassman, M. Manasse, and G. Zweig. Syntactic Clustering of the Web. In *Proceedings of the 6th WWW International Conference on World Wide Web*, pages 391–404, 1997.
[7] S.-H. Cha. Comprehensive Survey on Distance/Similarity Measures between Probability Density Functions. *International Journal of Mathematical Models and Methods in Applied Sciences*, 1(4):300–307, 2007.
[8] S.-H. Cha and S. Srihari. On Measuring the Distance between Histograms. *Pattern Recognition*, 35(6):1355–1370, 2002.
[9] M. Charikar. Similarity Estimation Techniques from Rounding Algorithms. In *Proceedings of the 34th ACM STOC Symposium on Theory Of Computing*, pages 380–388, 2002.

[10] S. Chaudhuri, V. Ganti, and R. Kaushik. A Primitive Operator for Similarity Joins in Data Cleaning. In *Proceedings of the 22nd IEEE International Conference on Data Engineering*, page 5, 2006.

[11] J. Dean and S. Ghemawat. Mapreduce: Simplified Data Processing on Large Clusters. In *Proceedings of the 6th USENIX OSDI Symposium on Operating System Design and Implementation*, pages 137–150, 2004.

[12] eHarmony Dating Site. http://www.eharmony.com.

[13] T. Elsayed, J. Lin, and D. Oard. Pairwise Document Similarity in Large Collections with MapReduce. In *Proceedings of the 46th HLT Meeting of the ACL on Human Language Technologies: Short Papers*, pages 265–268, 2008.

[14] S. Ghemawat, H. Gobioff, and S.-T. Leung. The Google File System. In *Proceedings of the 19th ACM SOSP Symposium on Operating Systems Principles*, pages 29–43, 2003.

[15] A. Gibbs and F. Su. On Choosing and Bounding Probability Metrics. *The International Statistical Review*, 70(3):419–435, 2002.

[16] K. Grauman and T. Darrell. Approximate Correspondences in High Dimensions. In *Proceedings of the 16th WWW International Conference on World Wide Web*, pages 505–512, 2006.

[17] M. Henzinger. Finding near-duplicate web pages: A large-scale evaluation of algorithms. In *Proceedings of the 29th ACM SIGIR Conference on Research and Development in Information Retrieval*, pages 284–291, 2006.

[18] P. Indyk and R. Motwani. Approximate Nearest Neighbors: Towards Removing the Curse of Dimensionality. In *Proceedings of the 19th ACM STOC Symposium on Theory Of Computing*, pages 604–613, 1998.

[19] C. Li, J. Lu, and Y. Lu. Efficient Merging and Filtering Algorithms for Approximate String Searches. In *Proceedings of the ICDE 42nd IEEE International Conference on Data Engineering*, pages 257–266, 2008.

[20] Y.-R. Lin, H. Sundaram, Y. Chi, J. Tatemura, and B. Tseng. Blog Community Discovery and Evolution Based on Mutual Awareness Expansion. In *Proceedings of the 6th IEEE/WIC/ACM WI International Conference on Web Intelligence*, pages 48–56, 2007.

[21] J. Linn and C. Dyer. *Data-Intensive Text Processing with MapReduce*. Synthesis Lectures on Human Language Technologies Series. Morgan & Claypool Publishers, 2010.

[22] A. Metwally, D. Agrawal, and A. El Abbadi. DETECTIVES: DETEcting Coalition hiT Inflation attacks in adVertising nEtworks Streams. In *Proceedings of the 16th WWW International Conference on World Wide Web*, pages 241–250, 2007.

[23] A. Metwally, F. Emekçi, D. Agrawal, and A. El Abbadi. SLEUTH: Single-pubLisher attack dEtection Using correlaTion Hunting. *Proceedings of the VLDB Endowment*, 1(2):1217–1228, 2008.

[24] A. Metwally and M. Paduano. Estimating the Number of Users Behind IP Addresses for Combating Abusive Traffic. In *Proceedings of the 17th ACM SIGKDD International Conference on Knowledge Discovery and Data Mining*, pages 249–257, 2011.

[25] C. Olston, B. Reed, U. Srivastava, R. Kumar, and A. Tomkins. Pig Latin: A Not-So-Foreign Language for Data Processing. In *Proceedings of the 28th ACM SIGMOD International Conference on Management of Data*, pages 1099–1110, 2008.

[26] R. Pike, S. Dorward, R. Griesemer, and S. Quinlan. Interpreting the Data: Parallel Analysis with Sawzall. *Scientific Programming*, 13(4):277–298, October 2005.

[27] J. Ruan and W. Zhang. An Efficient Spectral Algorithm for Network Community Discovery and Its Applications to Biological and Social Networks. In *Proceedings of the 7th IEEE ICDM International Conference on Data Mining*, pages 643–648, 2007.

[28] Y. Rubner, C. Tomasi, and L. Guibas. A Metric for Distributions with Applications to Image Databases. In *Proceedings of the 6th IEEE ICCV International Conference on Computer Vision*, pages 59–66, 1998.

[29] S. Sarawagi and A. Kirpal. Efficient Set Joins on Similarity Predicates. In *Proceedings of the 24th ACM SIGMOD International Conference on Management of Data*, pages 743–754, 2004.

[30] V. Satuluri and S. Parthasarathy. Scalable Graph Clustering Using Stochastic Flows: Applications to Community Discovery. In *Proceedings of the 15th ACM SIGKDD International Conference on Knowledge Discovery and Data Mining*, pages 737–746, 2009.

[31] R. Stanley. *Enumerative Combinatorics*, volume 1. Cambridge University Press, 2002.

[32] A. Thusoo, J. Sarma, N. Jain, Z. Shao, P. Chakka, S. Anthony, H. Liu, P. Wyckoff, and R. Murthy. Hive - A Warehousing Solution Over a Map-Reduce Framework. *Proceedings of the VLDB Endowment*, 2(2):1626–1629, August 2009.

[33] R. Vernica, M. Carey, and C. Li. Efficient Parallel Set-Similarity Joins Using MapReduce. In *Proceedings of the 30th ACM SIGMOD International Conference on Management of Data*, pages 495–506, 2010.

[34] C. Xiao, W. Wang, X. Lin, and J. Yu. Efficient Similarity Joins for Near Duplicate Detection. In *Proceedings of the 18th WWW International Conference on World Wide Web*, pages 131–140, 2008.

[35] Y. Yu, M. Isard, D. Fetterly, M. Budiu, U. Erlingsson, P. Gunda, and J. Currey. DryadLINQ: A System for General-Purpose Distributed Data-Parallel Computing Using a High-Level Language. In *Proceedings of the 8th USENIX OSDI Conference on Operating Systems Design and Implementation*, pages 1–14, 2008.

[36] H. Zhang, C. Giles, H. Foley, and J. Yen. Probabilistic Community Discovery Using Hierarchical Latent Gaussian Mixture Model. In *Proceedings of the 22nd AAAI National Conference on Artificial Intelligence*, pages 663–668, 2007.